\documentclass[aps,pre,amsmath,amssymb,reprint,twocolumn]{revtex4-2}
\usepackage{amsfonts,amsthm,amsmath,amssymb,graphicx,color}
\usepackage{mathrsfs}
\usepackage{hyperref}
\usepackage{float}
\usepackage{subcaption}
\usepackage{hhline}
\usepackage{bbold}

\begin{document}
\title{Monotonicity of eigenstate thermalization hypothesis in two-dimensional systems}
\author{Nilakash Sorokhaibam}
\email{phy\_sns@tezu.ernet.in}
\author{Anjan Daimari}
\affiliation{Department of Physics, Tezpur University, Tezpur 784028, Assam, India.}

\begin{abstract}
We study numerically the enveloping $f$-function of the fluctuation term in eigenstate thermalization hypothesis (ETH) statement. We concentrate on the energy (or entropy) dependence of this function in two-dimensional systems. Our numerical results show that it is, in general, a monotonically increasing function of the entropy. This is in agreement with the general expectation that fluctuations increase with increasing entropy. We show that the $f$-function locally flattens with increasing system-size. The flattening rate is directly proportional to the system size. We also show that the flattening rate is directly proportional to the particle number for systems of same spatial size. This variation of the $f$-function is important for physics at subleading order of the system-size. So, it is relevant for intermediate-size systems (upto a few hundred qubits) which are experimentally accessible. One exception we found is that the $f$-function of the order parameter of a thermal phase transition defy the monotonic behaviour.
\end{abstract}

\maketitle

\section{Introduction and Summary}
\label{intro_sec}

Quantum chaos and the related topic of thermalization in quantum many-body systems have been subjects of intense study for the past few decades \cite{Polkovnikov:2010yn,DAlessio:2015qtq,Gogolin:2016hwy}. Many-body quantum chaotic systems are expected to thermalize. It is widely believed that this is how (quantum) statistical mechanics arises in many-body systems.
On the other hand, integrable systems have been observed to avoid thermalization experimentally as well as numerically \cite{kinoshita2006quantum,2009PhRvL.103j0403R}.

One of the most important theoretical advances in this direction is the discovery of eigenstate thermalization hypothesis (ETH) \cite{PhysRevA.43.2046,Srednicki:1994mfb}. In quantum mechanics, we can keep track of only certain few-body observables which can be easily measured or calculated. ETH gives an ansatz of the matrix elements of such observables in the energy eigenbasis. It has been shown to be true for various chaotic models, see for examples \cite{Rigol_2010,Santos_20102,Steinigeweg_2013,Steinigeweg_2014,Kim:2014jfl,DAlessio:2015qtq,Fratus_2015,Mondaini_2016}. ETH (at least Srednicki's version) is motivated from Berry's conjecture \cite{Berry_1977,DAlessio:2015qtq}. Berry's conjecture states that an energy eigenstate of quantum chaotic system is observationally equivalent to a microcanonical ensemble. ETH statement can be considered as Berry's conjecture with precise definition of the fluctuations.

Consider a non-fermionic and hermitian operator $\mathcal{O}$ which represents one such observable. ETH states that the matrix elements $\mathcal{O}_{mn}$ in energy eigenbasis is \cite{DAlessio:2015qtq}
\begin{equation}
\langle m|\mathcal{O}|n\rangle\equiv \mathcal{O}_{mn}=\mathcal{O}(\bar{E})\delta_{mn}+e^{-S(\bar{E})/2} f(\bar{E},\omega)R_{mn}\
\label{ETH}
\end{equation}
$|m\rangle$ and $|n\rangle$ are energy eigenstates with energy eigenvalues $E_m$ and $E_n$ respectively. $\bar{E}=(E_m+E_n)/2$ and $\omega=E_m-E_n$ are the mean and the difference of $E_m$ and $E_n$. $S(\bar{E})$ is the entropy at energy $\bar{E}$. $\mathcal{O}(\bar{E})$ and $f(\bar{E},\omega)$ are smooth functions of their arguments. So, $\mathcal{O}(\bar{E})$ is equal to the expectation value of $\mathcal{O}$ in the microcanonical ensemble of energy $\bar{E}$. $R_{mn}$ are pseudo-random complex variables with zero mean and unit variance. The operator $\mathcal{O}$ is hermitian so
\begin{equation}
R_{nm}=R^*_{mn}, \; f(\bar{E},-\omega)=f(\bar{E},\omega)^*.
\end{equation}
For a clean presentation, we will consider $f(\bar{E},\omega)$ to be real and positive. All of our results hold true for $|f(\bar{E},\omega)|$ when $f(\bar{E},\omega)$ is a complex function.

In this paper, we study the $\bar{E}$-dependence of the $f$-function. We will write the $f$-function as $f(\bar{E})$ for a fixed $\omega$ when the $\omega$-dependence is not important. We will work with $f(\bar{E})$ as a function of the effective temperature $T(\bar{E})$ or effective inverse temperature $\beta(\bar{E})$ or entropy $S(\bar{E})$. With a slight abuse of notation, we will represent these new functions as $f(T)$, $f(\beta)$ or $f(S)$. Earlier work \cite{sorokhaibam2025quantumchaosarrowtime} in 1-dimensional systems suggest that this function is, in general, a monotonically increasing function of $S$ or $|T|$. The general motivation for this monotonic behaviour is that we expect the fluctuations to increase with increasing $S$ or $|T|$. We particularly meant the ``fluctuations within each eigenstate" (see eqn (210) in \cite{DAlessio:2015qtq})
\begin{eqnarray}
\delta \mathcal{O}^2_n&=&\langle n|\mathcal{O}^2|n\rangle-\langle n|\mathcal{O}|n\rangle^2\nonumber\\
&=&\int_{-\infty}^{\infty}d\omega e^{\beta\omega/2} f(E_n,\omega)^2\left[1+\left.\frac{\partial f(\bar{E},\omega)/\partial\bar{E}}{f(\bar{E},\omega)}\right|_{\bar{E}=E_n}\right.\nonumber\\
&&\hspace{3.8cm} \left.+\frac{3\omega^2 S''(\bar{E})|_{\bar{E}=E_n}}{8}\right].\
\label{fluctuations}
\end{eqnarray}
We have included $3\omega^2 S''(\bar{E})/8$ because it can compete with the $f'(\bar{E})/f(\bar{E})$ term. Compared to $1$, these terms are suppressed by powers of system-size because $\bar{E}$ and $S$ are extensive quantities. Consider the leading integrand. The entropic factor $e^{\beta\omega/2}$ is a decreasing function of $S$ or $|T|$ for a fixed $\omega$. So, we expect the $f$-function to be a fast increasing function of $S$ or $|T|$ because the overall fluctuation $\delta \mathcal{O}^2_n$ is an increasing function of $S(E_n)$ or $|T(E_n)|$. We are not considering the ``fluctuations from energy fluctuations" (see eqn (211) in \cite{DAlessio:2015qtq}) because we are interested in the ``fluctuation within each eigenstate" in the spirit of ETH and Berry's conjecture.

We work with two different models, viz. hard-core bosons (HCB) on a two-dimensional square lattice and transverse field Ising model (TFIM) on a two-dimensional square lattice, both with open boundary condition. The primary task is to calculate $f(\bar{E},\omega)$. Then the $f$-function can be studied as a function of $S$ or $|T|$. TFIM in two dimensions exhibits a thermal phase transition for small values of the transverse magnetic field.

Our numerical results show that
\begin{enumerate}
\item The $f$-function is, in general, a monotonically increasing function of entropy $S(\bar{E})$ or magnitude of the temperature $|T(\bar{E})|$.
\item We find that $f(\bar{E})$ flattens locally as $A^1$ as the system-size $A$ increases,
\begin{gather}
\frac{f'(\bar{E})}{f(\bar{E})}\;\xrightarrow{A\to\infty} \;0, \qquad A\;\frac{f'(\bar{E})}{f(\bar{E})}\; = \; \text{a constant.}
\label{ETHmono}
\end{gather}
In case of a system with conserved particle number, the linear particle density $N/\sqrt{A}$ is kept fixed.
\item In systems with particle number conservation, we find that $f(\bar{E})$ flattens locally as $N^1$ where $N$ is the number of particles.
\begin{gather}
\qquad N\;\frac{f'(\bar{E})}{f(\bar{E})}\; = \; \text{a constant}, \quad \text{fixed} \, A.
\label{ETHmono_N_ddim}
\end{gather}
\item The $f$-function of the order parameter of the thermal phase transition defy the monotonic behaviour.
\end{enumerate}
The requirement that the linear density $N/\sqrt{A}$ should be kept fixed is surprising. Normally we take a system with particle number conservation to the thermodynamic limit by keeping the particle density $N/A$ fixed. Despite this, we believe that the flattening of the $f(\bar{E})$ is due to the stretching of this function as the system size grows. We are considering systems with only local interactions. For these systems, the energy is an extensive quantity. But we emphasize that it is not a simple rescaling of the $f(\bar{E})$ function. We do not know the precise transformation of $f(\bar{E})$ as the system size increases. On the other hand, the role of the extensive nature of $\bar{E}$ is further supported by the observation that $f(T)$ or $f(\beta)$ does not flatten as the system size increases, because $T$ or $\beta$ are intensive quantities. In a general $d$-dimensional system, we expect that the flattening rate is proportional to the volume $V$ of the system.

From (\ref{ETHmono}), $f'(\bar{E})/f(\bar{E})$ is proportional to $1/A$, meaning suppression by the system size. So, the variation of $f(E)$ is not relevant for macroscopic systems. But, it is relevant for intermediate-size systems (upto a few hundred qubits) which are experimentally accessible (as in \cite{kinoshita2006quantum}, but with non-integrable interactions). The $f'(\bar{E})/f(\bar{E})$ term in (\ref{fluctuations}) is expected to be positive because of the monotonic behaviour. It is opposed by the $\omega^2 S''(\bar{E})$ term. $S''(\bar{E})=-C/\beta^2$ is negative because we expect our systems to have positive heat capacity $C$. So, the $f'(\bar{E})/f(\bar{E})$ term will dominate the $\omega^2 S''(\bar{E})$ at low frequency.


We work with the reduced Planck constant $\hbar=1$ and the Boltzmann constant $k_B=1$. In section \ref{mnm}, we describe the models under consideration in details. We also briefly explain the numerical techniques. The numerical results are presented in section \ref{results}. Section \ref{conclusions} is conclusions.

\section{Models and method}
\label{mnm}
We work with two different many-body quantum chaotic systems with open boundary condition. The first one is hard-core bosons on a square lattice with nearest neighbour interaction. This model is chaotic at high temperature away from the edges of the energy spectrum. Quantum chaos, thermalization and ETH in this model has been studied in \cite{Rigol_2008nature}. The one-dimensional version of this model with next-to-nearest neighbour interaction \cite{Rigol_2010,Santos_20102} is also chaotic and has been the subject of intense study for quantum chaos and ETH.

The Hamiltonian of the two-dimensional hard-core boson (HCB) system is
\begin{equation}
H_{B}=-\sum_{\langle i,j\rangle}\left(b^{\dagger}_ib_j+b^{\dagger}_jb_i\right)+U\sum_{\langle i,j\rangle}n_in_j.
\label{Hhcbosons}
\end{equation}
$\langle i,j\rangle$ denotes all nearest-neighbour sites. The first term is the kinetic term and the second term in the nearest-neighbour interaction term where $n_i=b^{\dagger}_ib_i$ is the occupation number at site $i$. The boson number is a conserved quantity. We work in a fixed particle sector of this Hamiltonian. We fix $U=0.1$ for which the system is chaotic.

We work with systems of size $4\times 4$ lattice sites with 4 bosons, $4\times 5$ lattice sites with 4, 5 and 6 bosons, $5\times 5$ lattice sites with 4 and 5 bosons, $3\times 5$ lattice sites with 5 bosons, $3\times 6$ lattice sites with 5, 6 and 7 bosons, and $3\times 7$ lattice sites with 6 bosons. The dimensions of the Fock space ranges from $1820$ upto $54264$. The Hamiltonian has discrete symmetries under $180^{\circ}$ rotation operation $\mathcal{R}_{\pi}$ and reflection operation about the x-axis $\mathcal{P}_x$ (reflection about the y-axis $\mathcal{P}_y=\mathcal{R}_{\pi}\mathcal{P}_x$) \cite{Sandvik_2010}. But we will not resolve these symmetries because it is not necessary for ETH analysis and it is well-known that this system is chaotic.

The Hamiltonian matrix is diagonalized using exact diagonalization. We considered two observables, namely, the kinetic energy term (an extensive operator) and the occupation number of a single site (an intensive operator).
\begin{gather}
\mathcal{O}_{B1}=-\sum_{\langle i,j\rangle}\left(b^{\dagger}_ib_j++b^{\dagger}_jb_i\right), \qquad \mathcal{O}_{B2}=n_i\
\end{gather}
The single site is chosen to be somewhere in the middle of the lattice so that the boundary effect is reduced. For example, site $(2,2)$ in the $4\times 4$ lattice, $(2,3)$ in the $4\times 5$ lattice and $(3,3)$ in the $5\times 5$ lattice, etc.

We calculate the above operators in the energy eigenbasis. The matrix elements lying within a small frequency window $d\omega$ centred around a fixed $\omega$ are identified. So, it is a collection of matrix elements with various possible values of $\bar{E}$ but with fixed values of $\omega$. Then we calculate the running average of the square of these matrix elements lying within a microcanonical energy window $d\bar{E}$. The unit variance of $R_{mn}$ implies that the calculated average is $e^{-S(\bar{E})} f(\bar{E},\omega)^2$, from which we can extract $f(\bar{E},\omega)$ easily. The microcanonical energy window $d\bar{E}$ is needed for the calculation of the running averages as well as for the calculation of the entropy $S(\bar{E})$. As in \cite{sorokhaibam2025quantumchaosarrowtime}, we choose $d\bar{E}$ in the most conservative manner so that the monotonic behaviour of the $f$-function remains robust. For example, the choice of $d\bar{E}$ is such that, even if we use the canonical entropy instead of the microcanonical entropy, the ETH-monotonic behaviour of the $f$-function remains unchanged. Actually the use of the canonical entropy reinforces the ETH-monotonic behaviour of the $f$-function. The $f$-function becomes a faster monotonically increasing function of $|T(\bar{E})|$. After calculating $f(\bar{E},\omega)$, we calculate $f'(\bar{E})/f(\bar{E})$. We then calculate the slope of $f'(\bar{E})/f(\bar{E})$ in an interval of $\bar{E}$ away from the edges of the spectrum to study the local flattening rate of $f(\bar{E})$.

The other system we are considering is the two-dimensional transverse field Ising model (TFIM) \cite{Fratus_2015,Mondaini_2016}. The Hamiltonian is
\begin{equation}
H_{S}=-\sum_{\langle i,j\rangle} \sigma_i^z\sigma_j^z-g\sum_i\sigma_i^x
\label{Htfim}
\end{equation}
$\sigma$'s are the Pauli matrices. The first term is the interaction term and the second term is the transverse magnetic field term. This system in the thermodynamic limit undergoes a quantum phase transition from the ferromagnetic phase to the disordered phase as the strength of the transverse magnetic field $g$ increases. The critical value of the field strength is $g_c \simeq 3.044$. The order parameter is the total magnetization $\mathcal{M}_z=\sum_i\sigma_i^z$. So for $g<g_c$, the system exhibits a second order thermal phase transition. But for a finite system with $g<g_c$, it was found that most of the eigenstates are in the disordered phase \cite{Mondaini_2016}. The one-dimensional Ising model with next-to-nearest neighbour interaction has also been studied in \cite{Steinigeweg_2013}.

The operators under consideration are the total transverse magnetization $\mathcal{O}_{S1}$, and the order parameter $\mathcal{M}_z$.
\begin{gather}
\mathcal{O}_{S1}=\sum_i\sigma_i^x, \qquad \mathcal{M}_z=\sum_i\sigma_i^z\
\end{gather}
We choose $g=1$ to study the monotonic behaviour of the $f$-functions of these operators. We choose this value of $g$ for two reasons. First, we are particularly interested in the monotonic behaviour of the $f$-function of the order parameter across the thermal phase transition. Indeed our system is small, there is no sharp thermal phase transition and most of the eigenstates are in the disordered phase. But we still find interesting results. Second, high value of $g\geq 2$ leads to development of patterns in the off-diagonal matrix elements of observables, even after taking care of all the discrete symmetries. It does not appear to be a complete violation of ETH statement, but the patterns give rise to sharp jumps in $f(\bar{E})$. The verification of the ETH statement in \cite{Mondaini_2016} was done only for the diagonal elements.

Following the intriguing observation above, we briefly study the chaotic nature of the system by analysing the energy spectrum. The total spin is not a conserved quantity so we have to work with the entire Fock space. We consider even states under $\mathcal{P}_x$ operation and $\mathcal{R}_{\pi}$ rotation. The Hamiltonian further has a spin-flip $\mathbb{Z}_2$ symmetry under the operation $\prod_i\sigma_i^x$. For the calculation of $f(\bar{E},\omega)$, we break this symmetry explicitly by introducing a small longitudinal magnetic field $\epsilon$ as in \cite{Fratus_2015}. This symmetry breaking is accomplished by introducing the term $\epsilon \mathcal{M}_z$ in the Hamiltonian. We used $\epsilon=0.0005$. We choose this approach mainly because we want to examine $f(\bar{E},\omega)$ of the order parameter $\mathcal{M}_z$. But for the study of the chaotic nature of the system, we do not break the $Z_2$ symmetry. Instead, we consider only the even states under the spin flip operation. We consider system sizes of $3\times 4$ lattice sites, $3\times 5$ lattice sites, $4\times 4$ lattice sites and $3\times 7$ lattice sites. The Fock space dimensions are $1120$, $8640$, $16576$ and $66816$ respectively, after taking care of the symmetries under the operation $\mathcal{P}_x$ and $\mathcal{R}_{\pi}$. If we further consider only even states of the spin-flip operation for the study of chaotic nature, the dimension of the Fock space is $4320$ for the system with $3\times 5$ lattice sites. $\langle\mathcal{M}_z\rangle=0$ for these states.

The Mathematica codes used for the numerical calculations are available as ancillary files at the arXiv webpage of this paper https://arxiv.org/abs/2510.25711.

\section{Results}
\label{results}

\begin{figure}
\begin{center}
\includegraphics[width=1.\columnwidth]{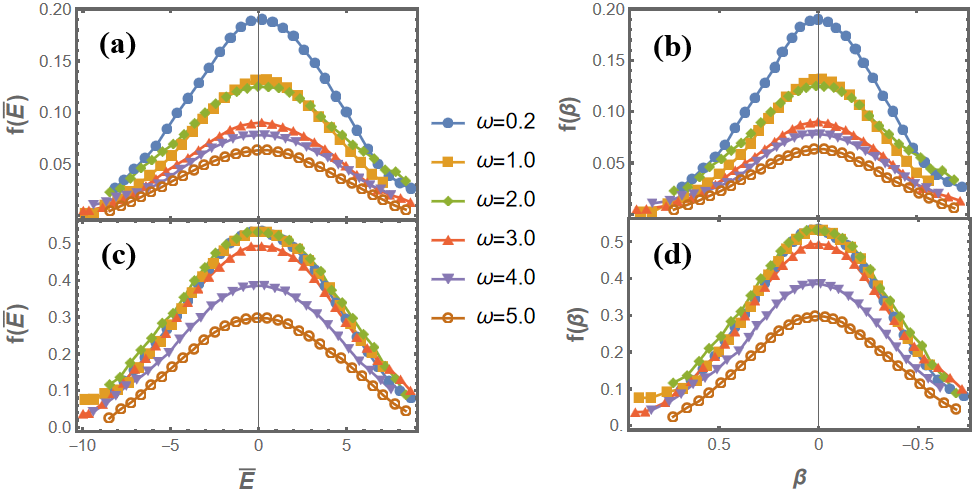}
\caption{\small Plots showing the monotonically increasing behaviour of $f(T)$ as a function of $T$. Top panels: Plots for the kinetic energy operator $\mathcal{O}_{B1}$. Bottom panels: Plots for the occupation number operator at a single site $\mathcal{O}_{B2}$.}
\label{HCB_KEocOrave_plots}
\end{center}
\end{figure}

\begin{figure}
\begin{center}
\includegraphics[width=1.\columnwidth]{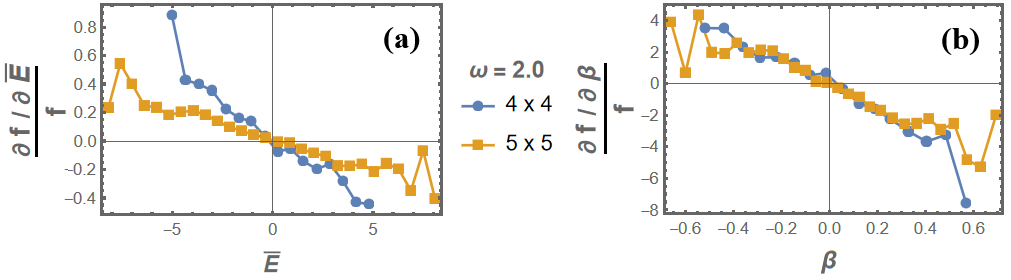}
\caption{\small Left panel: Plot of $f'(\bar{E})/f(\bar{E})$ for different system sizes showing that $f(\bar{E})$ flattens locally as the system size increases. Right panel: Plot of $f'(\beta)/f(\beta)$ for different system sizes showing that $f(\beta)$ does not flatten as the system size increases. Both panels are for the kinetic energy operator $\mathcal{O}_{B1}$.}
\label{HCB_KErave_dfEb_dfbeta_plots}
\end{center}
\end{figure}

We first present the results for the hard-core boson (HCB) system. We study $f(\bar{E},\omega)$ for $\omega\in [0.0,5.0]$. Figure \ref{HCB_KEocOrave_plots} shows the monotonically decreasing behaviour of $f(\beta)$ with varying $|\beta|$, for different fixed values of $\omega$. So, $f(T)$ is a monotonically increasing function of $|T|$. These plots are for the system of $5\times 5$ lattice sites with 5 bosons. For completeness, we have also plotted $f(\bar{E})$. The top row is for the kinetic energy operator $\mathcal{O}_{B1}$. The bottom row is for the occupation number operator at a single site $\mathcal{O}_{B2}$.

Figure \ref{HCB_KErave_dfEb_dfbeta_plots} are plots of $f'(\bar{E})/f(\bar{E})$ as a function of $\bar{E}$ (left panel) and $f'(\beta)/f(\beta)$ as a function of $\beta$  (right panel) for the two system sizes - $4\times 4$ lattice with $4$ bosons and $5\times 5$ lattice with $5$ bosons. The operator under consideration is $\mathcal{O}_{B1}$. The left panel shows that the $f(\bar{E})$ function locally flattens as the system size increases. On the other hand, we can see that $f(\beta)$ function does not flattens as the system size increases. This agrees with the fact that $\beta$ is an intensive quantity.

\begin{figure}
\begin{center}
\includegraphics[width=.75\columnwidth]{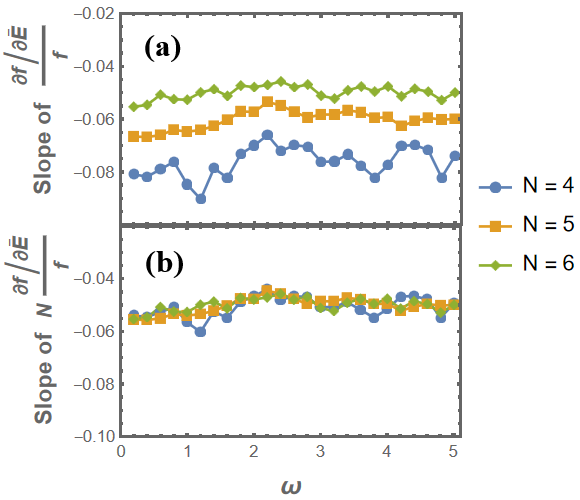}
\caption{\small For the kinetic energy operator $\mathcal{O}_{B1}$: (a) Slopes of $f'(\bar{E})/f(\bar{E})$ as a function of $\omega$ for different particle number, (b) The slopes overlap each other after rescaling with the total particle number $N$.}
\label{HCB_KE_Nfrate_plots}
\end{center}
\end{figure}

We present the plots of the slope of $f'(\bar{E})/f(\bar{E})$ for different boson numbers in Figure \ref{HCB_KE_Nfrate_plots}(a). The system size is $4\times 5$ lattice with 4, 5, and 6 bosons. The absolute value of the slope for the higher particle number is clearly smaller. Figure \ref{HCB_KE_Nfrate_plots}(b) is plots of the slope of $f'(\bar{E})/f(\bar{E})$ but rescaled with the boson number $N$. The plots for different boson numbers interweave each other now. Within numerical error, it is safe to say that the rescaled slopes are overlapping. This implies that the rate of flattening of $f(\bar{E})$ is directly proportional to the particle number $N$ for a fixed lattice size. We have similar results for $3 \times 6$ system with 5, 6 and 7 bosons, and $5 \times 5$ system with 4 and 5 bosons. We get similar results for the occupation number operator at a single site $\mathcal{O}_{B2}$.

\begin{figure}
\begin{center}
\includegraphics[width=.85\columnwidth]{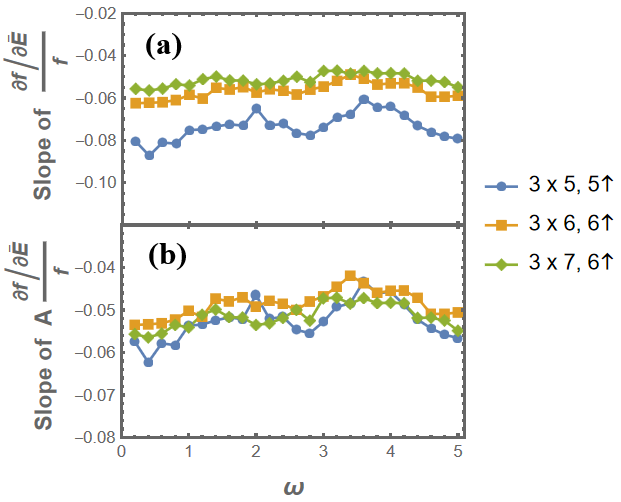}
\caption{\small For the kinetic energy operator $\mathcal{O}_{B1}$: (a) Slopes of $f'(\bar{E})/f(\bar{E})$ as a function of $\omega$ for different system sizes $A$, (b) The slopes rescaled with the system size $A$.}
\label{HCB_KE_Afrate_plots}
\end{center}
\end{figure}

\begin{figure}
\begin{center}
\includegraphics[width=.85\columnwidth]{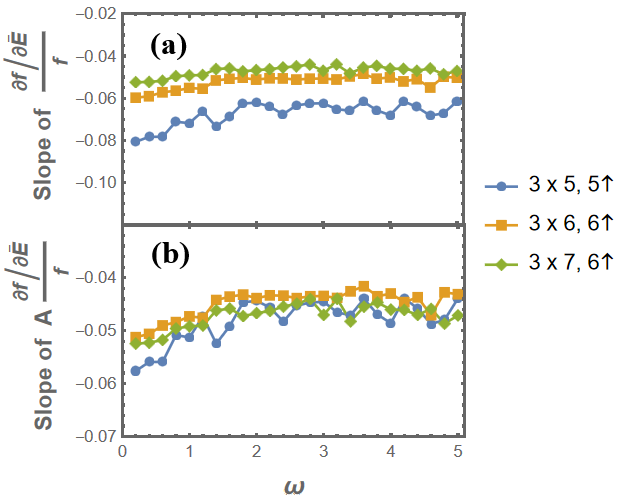}
\caption{\small For the occupation number of a single site $\mathcal{O}_{B2}$: (a) Slopes of $f'(\bar{E})/f(\bar{E})$ as a function of $\omega$ for different system sizes $A$, (b) The slopes rescaled with the system size $A$.}
\label{HCB_ocO_Afrate_plots}
\end{center}
\end{figure}

We use Figure \ref{HCB_KE_Afrate_plots} to find out the rate of local flattening of $f(\bar{E})$ as the system size increases. The flattening rate scales as the system-size. But what is most surprising is the observation that the linear particle density $N/\sqrt{A}$ should be fixed, instead of the particle density $N/A$. Figure \ref{HCB_KE_Afrate_plots}(a) are plots of the slope of $f'(\bar{E})/f(\bar{E})$ for the different system sizes considered. We present the result for $3\times 5$ system with 5 bosons (X), $3\times 6$ system with 6 bosons (Y) and $3\times 7$ system with 6 bosons (Z). The particle density $N/A$ are $0.33$ particles per site, $0.33$ particles per site and $0.29$ particles per site respectively. The linear particle density $N/\sqrt{A}$ are $1.29$ particles per site, $1.41$ particles per site and $1.30$ particles per site respectively. Figure \ref{HCB_KE_Afrate_plots}(b) are plots of the slope rescaled with the system size $A$. The rescaled slope of system X does not overlap/interweave with that of system Y. Instead, the overlap is between system X and system Z. This means that the flattening rate of $f(\bar{E})$ is proportional to $A$ for fixed linear particle density $N/\sqrt{A}$. We get similar results in Figure \ref{HCB_ocO_Afrate_plots} for the occupation number operator at a single site $\mathcal{O}_{B2}$. We have similar result for another group of systems - $4\times 4$ system with 4 bosons (X), $4\times 5$ system with 5 bosons (Y) and $5\times 5$ system with 5 bosons (Z). The particle density are $0.25$ particles per site, $0.25$ particles per site and $0.2$ particles per site respectively. The linear particle density are $1.00$ particles per site, $1.12$ particles per site and $1.00$ particles per site respectively. Overlapping of the rescaled slopes is again observed only for the two systems X and Z.

\begin{figure}
\begin{center}
\includegraphics[width=0.65\columnwidth]{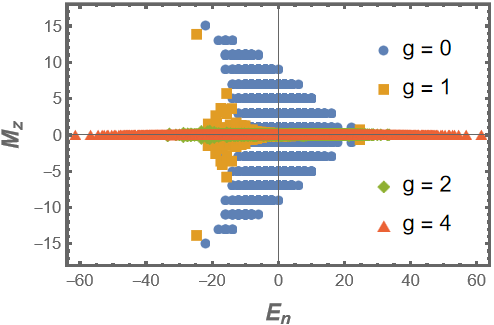}
\caption{\small Total magnetization $\langle M_z\rangle$ of energy eigenstates for different values of the transverse magnetic field $g$.}
\label{TFIM_magnetization_plots}
\end{center}
\end{figure}

We now present the results for the transverse field Ising model (TFIM). We study $f(\bar{E},\omega)$ for $\omega\in [0.0,25.0]$. Figure \ref{TFIM_magnetization_plots} is the plot of the total magnetization $\langle M_z\rangle$ of the eigenstates with different values of the transverse magnetic field $g$. The magnetization of all eigenstates rapidly decreases with increasing $g$.

\begin{figure}
\begin{center}
\includegraphics[width=1.\columnwidth]{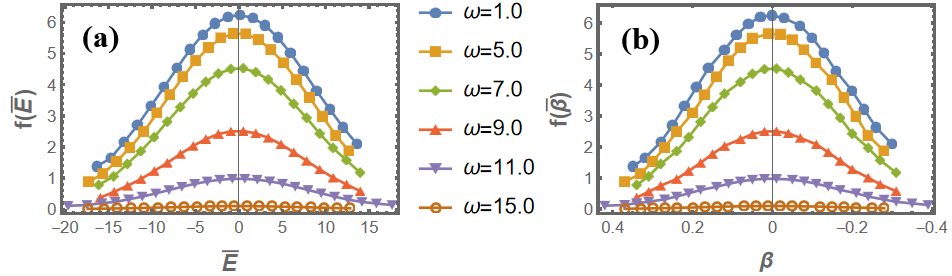}
\caption{\small $f(\bar{E})$ (left panel) and $f(\beta)$ (right panel) for different fixed $\omega$ for the total transverse magnetization $\mathcal{O}_{S1}$.}
\label{TFIM_KErave_Eb_beta_plots}
\end{center}
\end{figure}

As mentioned above, we work with $g=1$. Figure \ref{TFIM_KErave_Eb_beta_plots} are plots of $f(\bar{E})$ and $f(\beta)$ for the total transverse magnetization $\mathcal{O}_{S1}$. These plots show the monotonically decreasing behaviour of $f(\beta)$ as a function of $|\beta|$, so the $f$-function is a monotonically increasing function of $|T|$ or $S$.

\begin{figure}
\begin{center}
\includegraphics[width=.75\columnwidth]{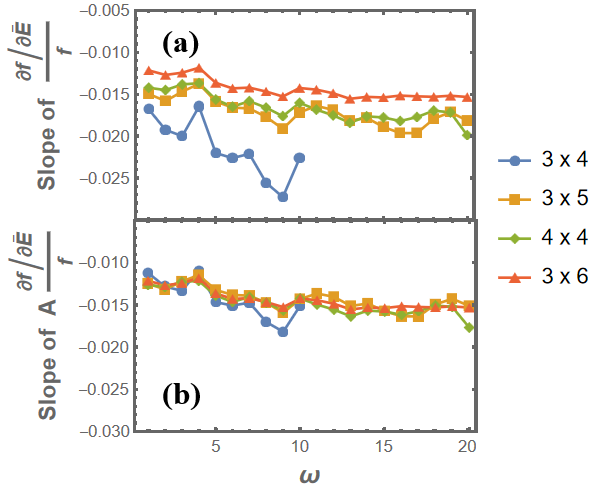}
\caption{\small For the transverse magnetization operator $\mathcal{O}_{S1}$: (a) Slopes of $f'(\bar{E})/f(\bar{E})$ as a function of $\omega$ for different system sizes $A$, (b) The slopes rescaled with the system size $A$.}
\label{TFIM_KE_frate_plots}
\end{center}
\end{figure}

Figure \ref{TFIM_KE_frate_plots} are plots of the slope of $f'(\bar{E})/f(\bar{E})$ for the total transverse magnetization $\mathcal{O}_{S1}$. It shows that the flattening rate of the $f(\bar{E})$ is directly proportional to the system-size $A$. After rescaling with $A^1$ factor, the slopes of $f'(\bar{E})/f(\bar{E})$ for different system-sizes at different values of $\omega$ interweave each other. Within numerical error, it is safe to say that the rescaled slopes overlap each other. We have compared four different system-sizes. The smallest system with $3\times 4$ lattice sites has only $1120$ energy eigenstates. So, we have not calculated $f(\bar{E},\omega)$ with $\omega>10.0$. Even with such a small dimensional Fock space, the rescaling law is nicely producible.

\begin{figure}
\begin{center}
\includegraphics[width=1.\columnwidth]{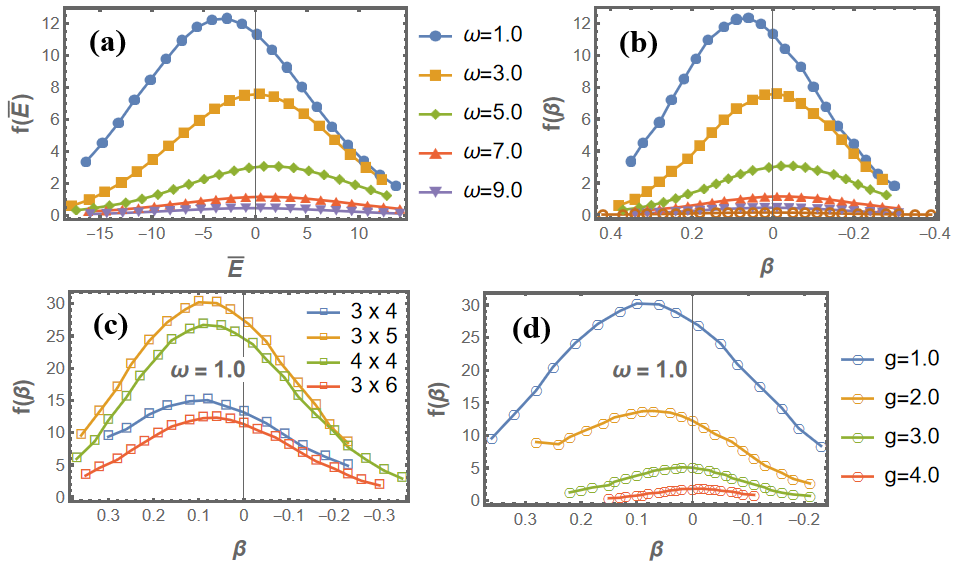}
\caption{\small For the order parameter $\mathcal{M}_{z}$: (a) plots of $f(\bar{E})$ for different values of $\omega$, (b) plots of $f(\beta)$ for different values of $\omega$, (c) plots of $f(\beta)$ with $\omega=1.0$ for different system sizes, (d) plots of $f(\beta)$ with $\omega=1.0$ for different values of transverse magnetic field $g$ for a fixed system size.}
\label{magOz_TFIM_plots}
\end{center}
\end{figure}

On the other hand, the $f(\bar{E})$ of the order parameter $\mathcal{M}_{z}$ is not perfectly a monotonically increasing function of $|T|$, see Figure \ref{magOz_TFIM_plots}(b). This is expected because the fluctuations is largest at the transition point of the second order phase transition. Moreover, the thermal phase transition is not at $\beta=0$. The thermal phase transition means a sharp discontinuity in the diagonal elements in the thermodynamic limit. The peak of $f(\bar{E})$ is farther away from $\beta=0$ for small $\omega$, meaning near the diagonal. Figure \ref{magOz_TFIM_plots}(c) is plots of $f(\beta)$ for different system sizes but with fixed transverse magnetic field $g=1$. The position of the peak does not change noticeably as the system size varies. Figure \ref{magOz_TFIM_plots}(d) is plots of $f(\beta)$ for different values of $g$ for the $3 \times 5$ system. It can be seen that the peak moves towards $\beta=0$ with increasing value of $g$. This is expected because, in a rough sense, more energy eigenstates move to the disordered phase as $g$ increases. We choose $\omega=1.0$ close to the diagonal for Figure \ref{magOz_TFIM_plots}(d) because the non-monotonic behaviour is more pronounced closer to the diagonal and, as we will soon see, the $f$-function starts having sharp discontinuities for large values of $g$ away from the diagonal. So, all the numerical results for $\mathcal{M}_z$ are expected. Actually we are surprised with the observation that the other observable $\mathcal{O}_{S1}$ still has the monotonic behaviour.

\begin{figure}
\begin{center}
\includegraphics[width=.6\columnwidth]{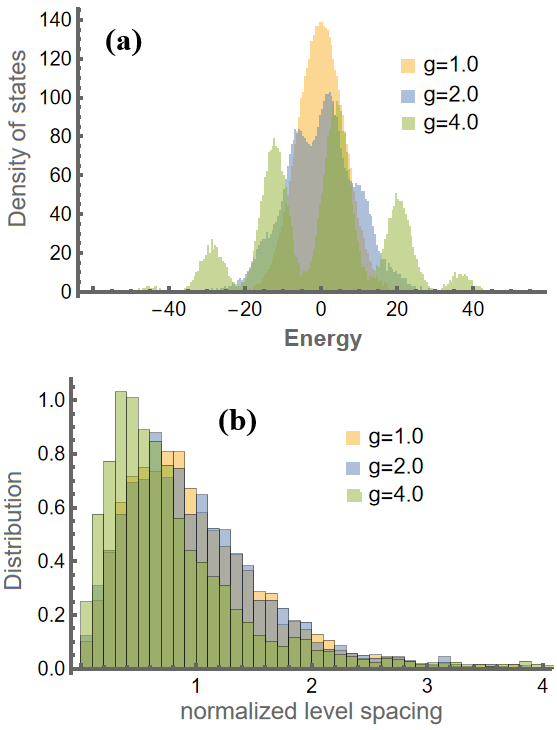}
\caption{\small (a) Density of states of TFIM for different values of the transverse magnetic field $g$. (b) Distribution of the level spacing of TFIM for different values of the transverse magnetic field $g$.}
\label{g124_spectrum}
\end{center}
\end{figure}

We also briefly studied the chaotic nature of TFIM with varying strength of the transverse magnetic field $g$. We consider the $3 \times 5$ system. We resolve all the three discrete symmetries. The eigenstates are even under all the discrete symmetries, including the spin-flip $\mathbb{Z}_2$ symmetry. Figure \ref{g124_spectrum}(a) is plots of the density of states for different values of $g$. It is clearly visible that the energy eigenvalues start clumping together into different sectors with increasing value of $g$. Figure \ref{g124_spectrum}(b) is plots of the distribution of level spacings. All the plots appear to approximately reproduce the Wigner-Dyson distribution expected for chaotic systems. But it is clearly visible that the extent of level-repulsion, which is a hallmark of chaotic systems, is decreasing with increasing $g$. We plan to undertake a more rigorous study of these observations in future. For the present paper, we have concentrated on the behaviour of the $f$-function.

\begin{figure}
\begin{center}
\includegraphics[width=1.\columnwidth]{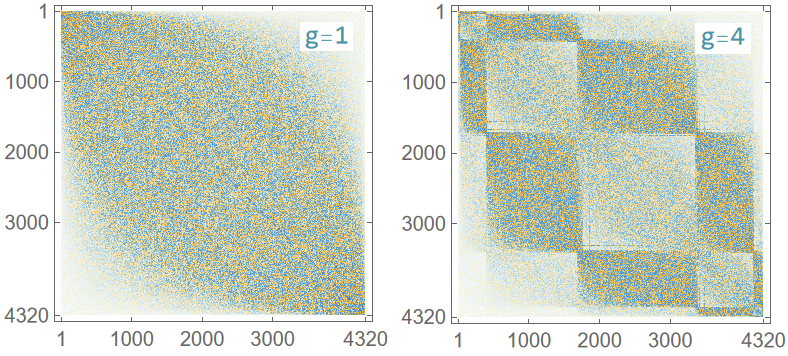}
\caption{\small Plots of the matrix elements of the operator $\mathcal{O}_{S1}$ in the energy eigenbasis for different values of the transverse magnetic field $g$. We used the function {\bf MatrixPlot} in Wolfram Mathematica. Vanishing values are whitish, negative values are bluish and positive values are reddish.}
\label{g10g40_KElist}
\end{center}
\end{figure}

Lastly, we present the full matrix structure of the operator $\mathcal{O}_{S1}$ in the energy eigenbasis. Figure \ref{g10g40_KElist} are plots of the matrix elements using the function {\bf MatrixPlot} in Wolfram Mathematica. Vanishing values are whitish, negative values are bluish and positive values are reddish. At $g=1$, the matrix elements are random as we expect from ETH statement (\ref{ETH}). For $g=4$, the matrix elements develop patterns. Actually we can see the pattern developing for $g\geq 2$. Such patterns give rise to sharp jumps in $f(\bar{E},\omega)$. $f(\bar{E},\omega)$ is expected to be a smooth function of its arguments. Validity of ETH for the diagonal elements has been studied in \cite{Mondaini_2016}. As we can see Figure in \ref{g10g40_KElist}, the matrix elements close to the diagonal appears to agree well with ETH statement. We also plan to have a closer look at these unexpected patterns in future. These patterns develop irrespective of the Fock space under consideration, they also appear in the full $2^A$-dimensional Fock space.

\section{Conclusions}
\label{conclusions}
In this paper, we studied the enveloping function $f(\bar{E},\omega)$ of the fluctuation term in ETH statement (\ref{ETH}). We concentrated on the $\bar{E}$-dependence for fixed values of $\omega$. General consideration dictates that this function should be a monotonically increasing function of entropy $S(\bar{E})$ or the magnitude of the effective temperature $|T(\bar{E})|$. We numerically examined this behaviour in two two-dimensional models, viz., hard-core boson (HCB) system and transverse field Ising model (TFIM). We found that observables in these systems, in general, possess the expected monotonic behaviour. An exception arises for the order parameter of the thermal phase transition in TFIM. It is expected because the fluctuations is largest at the transition point of the second order phase transition, away from $\beta=0$. The variation of the $f$-function is important for physics at subleading order of the system-size. So, it is relevant for intermediate-size systems (upto a few hundred qubits) which are experimentally accessible.

We found that $f(\bar{E})$ locally flattens, meaning $f'(\bar{E})/f(\bar{E}) \to 0$ as the system size increases. We found that the flattening rate is directly proportional to the system size $A$. The linear particle density $N/\sqrt{A}$ is fixed if the particle number is conserved. In general, we expect that the flattening rate is directly proportional to the volume $V$ for a d-dimensional system. Again, if the particle number $N$ is conserved, we also showed that $f(\bar{E})$ locally flattens with increasing particle number for a fixed system size. The flattening rate is directly proportional to $N$.

We are considering systems with only local interactions. For these systems, the energy is an extensive quantity. So, we believe that the local flattening of $f(\bar{E})$ is due to the extensive nature of $\bar{E}$ as the system size grows. But a simple rescaling of the $f$-function with the system size does not reproduce the correct $f(\bar{E})$. So, it appears that the flattening of the $f$-function is a more complicated transformation of the $f$-function in the $\bar{E}-\omega$ plane. This is further complication by the observation that the linear particle density $N/\sqrt{A}$ should be fixed, not the particle density $N/A$.

We also reported the intriguing observation in TFIM with increasing strength of the transverse magnetic field. The energy eigenvalues started clumping together into different sectors. Moreover, the matrix elements of the observables develop patterns which are not expected from ETH.

\begin{acknowledgments}
NS is fully supported by the Department of Science and Technology (Government of India) under the INSPIRE Faculty fellowship scheme IFA-20-PH-262. AD acknowledges the Department of Science and Technology - Anusandhan National Research Foundation (DST-ANRF, then DST-SERB) of the Government of India for providing Junior Research Fellowship under POWER Grant No. SPG/2022/000678.
\end{acknowledgments}

\bibliography{refs} 
\bibliographystyle{apsrev4-2}

\end{document}